\documentstyle[preprint,aps,12pt]{revtex}
\begin{document}

\title{Low Energy Behavior of the $^{7}$Be $(p,\gamma )^{8}$B
reaction}

\author{B.K.~Jennings }

\address{ TRIUMF, 4004 Wesbrook Mall, Vancouver, BC,
Canada, V6T 2A3}

\date{\today}
\maketitle

\begin{abstract}
Three related features of the astrophysical S-factor, $S_{17}$, for
low energies are an upturn as the energy of the proton goes to zero, a
pole at the proton separation energy and a long-ranged radial integral
peaking at approximately 40~fm. These features, particularly the last,
mean $S_{17}$ at threshold is largely determined by the asymptotic
normalization of the proton bound-state wave function. In this paper
we identify the pole contribution to the S-factor without any explicit
consideration of the asymptotic behavior of the bound-state
wave-function. This allows us to calculate, $S_{17}$ in terms of
purely short range integrals involving the two-body potential. Much of
the discussion is relevant to other radiative capture processes with
weakly bound final states.
\end{abstract}

The decay of $^{8}$B in the sun produces high-energy neutrinos that
are observed by earth based detectors. In fact, water detectors are
mainly sensitive these neutrinos. The $^{8}$B is produced in the
$^{7}$Be$(p,\gamma )^{8}$B reaction. In the sun this reaction is
peaked for protons with an energy of about 18~keV. The reader is
referred to Adelberger et al\cite{adl} for a review. Unfortunately the
low cross-sections make it highly unlikely that earth based
experiments will allow for measurements below 100~keV and even
measurements at that low an energy are very difficult. Thus the
experimental results have to be extrapolated from above 100~keV to
18~keV in order to be useful. This extrapolation would not seem like
much of a problem, except that $S_{17}$ has an upturn near threshold
and that upturn does not play an important role in the region
accessible to experiment. Thus the upturn has to be determined by
purely theoretical considerations --- a worrisome situation. In this
paper we will present a very simple derivation of the pole
contribution that causes the upturn. This new derivation does not
explicitly rely on the asymptotic behavior of the bound-state
wave-function. It does, however, show a remarkable and intimate
connection between the short distance part of the wave function and
the asymptotic normalization.  Furthermore, in addition to increasing
our confidence in our understanding of the low energy behavior of the
$S$-factor, it may provide a useful computational technique.

The long range of the radial integrals needed for determining the
$^{7}$Be$(p,\gamma )^{8}$B S-factor at low energies makes the
calculations more difficult then might be naively expected. For
example, shell model calculations are rarely valid in the extreme tail
region where the integrals are peaked. The long range of the integrals
can also make the calculation numerically difficult or time consuming,
even if the tail of the wave function is well determined.

On the other hand the long range of the integrals can be turned into
an advantage.  Since most of the contribution at the low energies
required in the solar calculations comes from the asymptotic region
outside the range of the nuclear force, the low energy $S$-function,
$S_{17}$, is determined mainly by the asymptotic normalization of the
bound-state wave-function and properties of Coulomb
wave-functions. Thus $S_{17}(0)$ is given once the asymptotic
normalization is known\cite{muk} without the need for additional
calculations. From ref.~\cite{jennings} one infers that
$S_{17}(0)=38A_{n}$ where $A_{n}$ is the asymptotic
normalization. This is accurate to about 1\%.

As has been recently pointed out, ref.~\cite{muk}, the asymptotic
normalization also determines the pole of the t-matrix for the elastic
scattering of protons on $^{7}$Be. Using this as a starting point we
will derive expressions for the asymptotic normalization and the
$S$-factor that depend only on the wave-function in the nuclear
interior. We start by writing the t-matrix as:
\begin{equation}
\label{eq1}
\langle k|T|k\rangle=\langle k|V|k\rangle+\langle k|VGV|k\rangle
\end{equation}
where $|k\rangle$ is a plane wave, $T$ is the t-matrix, $V$ is the potential
and $G$ is the full (not free) Greens function. Inserting the
eigenstates, $|\psi _{i}\rangle$, of the full Hamiltonian as a complete set
of states, we have:
\begin{equation}
\label{eq2}
\langle k|T|k\rangle=\langle k|V|k\rangle+\sum ^{\infty
}_{i=1}\frac{\left| \langle k|V|\psi _{i}\rangle\right|
^{2}}{E_{k}-E_{i}}
\end{equation}
where the $E_{i}$ are the eigen-energies of the full Hamiltonian and
$E_{k}$ is the energy corresponding to the plane wave $|k\rangle$. In
the continuum the sum becomes an integral. The t-matrix,
$\langle k|T|k\rangle$, has a pole at each of the bound-state
energies. The residues of these poles can be identified from
eq.~\ref{eq2}. For the $i$-th state it is $\left| \langle k|V|\psi
_{i}\rangle\right| ^{2}$.  Since the potential limits the radial
integral to relatively small values of $r$, this results shows that
the residue is determined purely by short-ranged parts of the wave
function; seemingly in contradiction with the claim of \cite{muk} that
the residue is determined by the asymptotic normalization of the
bound-state wave function. This apparent contradiction can be overcome
only if there is a relation between the short-range and long-range
parts of the wave-function. This is indeed that case as can be seen
using the equation:
\begin{equation}
\label{eq3}
\langle k|V|\psi _{i}\rangle=\langle k|H-H_o|\psi
_{i}\rangle=(E_{i}-E_{k})\langle k|\psi _{i}\rangle
\end{equation}
where $H$ is the full Hamiltonian and $H_{o}$ is free Hamiltonian.  In
deriving the last expression on the right we have used the fact that
$|k\rangle$ is an eigenstate of $H_o$. If we want $V$ to be just the
nuclear potential, excluding the Coulomb potential, then the Coulomb
potential must be included in $H_o$ and $|k\rangle$ becomes the
Coulomb distorted wave. This is useful in restricting the range of the
integral to range of the nuclear potential. A technique similar to
that leading to eq.~\ref{eq3} was used in ref.\cite{philpott} in
the context of R-matrix theory.

Rewriting the last equation we have:
\begin{equation}
\label{eq3b}
\langle k|\psi _{i}\rangle=\frac{\langle k|V|\psi
_{i}\rangle}{(E_{i}-E_{k})}
\end{equation}
The right hand side of this equation has an explicit pole at $E_{i} =
E_{k}$. On the left hand side the pole is not explicit but is due to
the radial integral diverging at $k=\pm i \kappa = \pm i \sqrt{2 m
|E_i|}$. Near the pole the bound-state wave-function can be replaced
by its asymptotic form for large $r$, $A_n \exp[-\kappa r]/r$, since the
integral is dominated by large $r$. For simplicity we have ignored the
Coulomb potential. With the asymptotic approximation for $|\psi
_{i}\rangle$, $\langle k|\psi _{i}\rangle$ becomes:
\begin{eqnarray}
   \langle k|\psi _{i}\rangle&\approx&
                    \frac{1}{(2\pi)^{3/2}}\int\exp[-ik\cdot r]A_n
                    \exp[-\kappa r]/rd^3r\nonumber\\ &\approx&
                    \frac{2 A_n}{(2 \pi)^{1/2} (\kappa^2+k^2)}
                    =\frac{A_n}{ m (2 \pi)^{1/2} (-E_i+E_k)},
\end{eqnarray} 
where $E_i$ is the energy of the bound state and is negative. The
integral in the first line can be carried out analytically and the
leading contribution to the integral for $k$ near $\pm i \kappa$ is
given in the second line. Thus, in this limit $\langle k|\psi
_{i}\rangle\propto \frac{A_{n}}{E_i-E_k}$ and we have $A_n=-(2
\pi)^{1/2}m \langle i\kappa | V | \psi_i\rangle$. Near $k=\pm
i\kappa$, $E_k$ is negative and as noted above for a bound state $E_i$
is also negative. Combining eq.~\ref{eq2} and eq.~\ref{eq3} we get:
\begin{eqnarray}
\label{eq3a}
\langle k|T|k\rangle&=&\langle k|V|k\rangle+\sum ^{\infty
}_{i=1}(E_{k}-E_{i})\left| \langle k|\psi _{i}\rangle\right| ^{2}
\nonumber\\ &\approx & \frac{1}{2 \pi m^2}\left| \frac{A_n}{E_k-E_i}
\right|^2
\end{eqnarray}
where the second line is valid near the pole. The residue of the pole
in the t-matrix is now seen to be proportional to $|A_{n}|^{2}$ or the
square of the absolute value of the asymptotic normalization. This is
in agreement with ref.~\cite{muk}.

The key to the above result is Eq.~\ref{eq3b}, which is just the
momentum-space bound-state Lippmann-Schwinger equation. It has the
amazing property of relating short and long-range properties of the
wave-function --- an integral over the interior of the nucleus to the
asymptotic normalization. In fact, if we take $|k\rangle$ to be a
Coulomb distorted wave function rather than a plane wave then the $V$
in eq.~\ref{eq3b} is just the nuclear potential and the range of the
integral is the range of the nuclear potential. A similar technique
was used in ref.~\cite{muk2} for proton and neutron emission.
Eq.~\ref{eq3b} can easily be extended to the case of a two body
potential which is needed for realistic calculations. Again the
integrals will be short-ranged but will contain two-body matrix
elements.

As has been stressed by Mukhamedzhanov and collaborators\cite{muk},
the asymptotic normalization is sufficient to determine $S_{17}$ near
threshold. By explicitly identifying the pole, eq.~\ref{eq3b} leads to
an expression for the asymptotic normalization in terms of an integral
over the interior of the nucleus. This may be useful in theoretically
calculating the asymptotic normalization and hence $S_{17}$ .  For
example, the shell model can give a good approximation to the wave
function in the interior while it typically does not do a good job in
describing the tail of the wave function. By construction the
scattering wave in eq.~\ref{eq3b} is the eigenstate of $H_o$ and thus
independent of the nuclear potential. Thus we can evaluate $\langle
k|V|\psi _{i}\rangle$ with $|\psi _{i}\rangle$ obtained from the shell
model and use eq.~\ref{eq3b} to determine the asymptotic
normalization.

Eq.~\ref{eq3b} is of use in determining $S_{17}$ only in the low
energy region where the integrals are long ranged and the S-factor is
determined mainly by the asymptotic normalization. It is, however,
possible to derive an expression that is short-ranged and valid for
all energies. This is done by identifying the singularity in the full
expression for $S_{17}$. At low energies the reaction is predominately
E1 and we take the dipole approximation. Starting with the matrix
element of the dipole operator and considering first the case of a
single particle in a one-body potential, the matrix element can be
written:
\begin{equation}
\label{eq4}
M=\langle\psi _{i}(k)|r|\psi _{f}\rangle
\end{equation}
where $\psi _{i}(k)$ is the fully-distorted wave in the initial
channel (contrast with eq,~\ref{eq3b}) and $\psi _{f}$ is the
wave-function of the final-state bound proton. This matrix element has
a second order pole\cite{jennings} as the energy of the scattering
state goes to the unphysical energy corresponding to the bound
state. To explicitly identify the singularity we proceed as follows:
\begin{equation}
\label{eq5}
M=\frac{\langle\psi _{i}(k)|Hr-rH|\psi
_{f}\rangle}{E_{k}-E_{f}}=-\frac{\langle\psi _{i}(k)|\frac{\nabla
}{m}|\psi _{f}\rangle}{E_{k}-E_{f}}
\end{equation}
where $E_{k}$ and $E_{f}$ are the energies of the initial and final
states respectively and we have assumed a local potential. Since $V$
is local it commutes with $r$. Thus $H r-r H=[H,r]$ reduces to the
commutator of $r$ with the kinetic energy. This gives the result shown
above.  Repeating this procedure of introducing the commutator with
$H$ gives (again assuming a local potential):
\begin{equation}
\label{eq6}
M=-\frac{\langle\psi _{i}(k)|H\frac{\nabla }{m}-\frac{\nabla
}{m}H|\psi _{f}\rangle}{\left( E_{k}-E_{f}\right)
^{2}}=\frac{\langle\psi _{i}(k)|\frac{\nabla V}{m}|\psi
_{f}\rangle}{\left( E_{k}-E_{f}\right) ^{2}}
\end{equation}
The second order pole is now explicitly shown. This equation differs
from eq.~\ref{eq3b} in having a higher power divergence and in using
the fully interacting wave-function for both the initial and finial
state.

The integral in the matrix element on the right hand side of
eq.~\ref{eq6} is restricted to the range of the potential. For a
proton the potential has a $1/r$ tail from the Coulomb potential so
the integrand will fall off a factor of $r^{3}$ faster then that in
eq.~\ref{eq4}. The integral in eq.~\ref{eq4} typically peaks at
40~fm\cite{jennings} with non-negligible contributions coming from
beyond 100fm. Repeating the S-factor calculations of
ref.~\cite{jennings} with eq.~\ref{eq6} indicates that the integrals
can now be cut off at about 50--60fm at threshold. While this is a
definite improvement, the integrals still extend far into the tail
region. As we will see shortly, when a two-body potential is used this
changes dramatically.

The physical case, of course, does not involve a one-body potential
but rather a two or more body interaction. For a many-body system the
dipole interaction is $\sum ^{A}_{i=1}e_{i}r_{i}$ and for a local
two-body interaction the potential is $\frac{1}{2}\sum
_{j,k}V(r_{j}-r_{k})$. Repeating the above procedure of introducing
the second order commutator we have :
\begin{equation}
\label{eq7}
M=\frac{\langle\psi _{i}(k)|\sum _{j,k}\frac{e_{j}\nabla
_{j}V(r_{j}-r_{k})}{m_{j}}|\psi _{f}\rangle}{\left( E_{k}-E_{f}\right)
^{2}}
\end{equation}
 For a two-body interaction Newton's third law (conservation of
momentum) gives us $\nabla _{j}V(r_{j}-r_{k})=-\nabla
_{k}V(r_{k}-r_{j})$. Thus if $e_{1}/m_{1}=e_{2}/m_{2}$, the \(
j=1$,$k=2$ term cancels the $j=2$,$k=1$ term. Applying this result for
all like pairs the matrix element reduces to :
\begin{equation}
\label{eq8}
M=\frac{e_{p}}{m_{p}}\frac{\langle\psi _{i}(k)|\sum _{j,k}\nabla
_{j}V(r_{j}-r_{k})|\psi _{f}\rangle}{\left( E_{k}-E_{f}\right) ^{2}}
\end{equation}
where $j$ is restricted to protons and $k$ to neutrons, \( e_{p}$ is
the charge of the proton and $m_{p}$ is its mass. Thus the matrix
element depends only on the neutron-proton potential and not on the
proton-proton potential.

In one-body models (see Christy and Duck\cite{christy}) a factor of
$\frac{e_{p}}{m_{p}}-\frac{e_{7}}{m_{7}}$, where $e_{7}$ and $m_{7}$
are the charge and mass of the target nucleus, is included to take
into account that the photon can couple either to the proton or to the
nucleus. Ignoring the binding energy this factor is just \(
\frac{e_{p}}{m_{p}}\frac{N}{A}$.  The factor of $e_{p}/m_{p}$ is
explicitly present in eq.~\ref{eq8}. The factor of $N/A$ can be
recovered if the sum over the neutrons can be replaced by $N/A$ times a
sum over all nucleons. This would be possible if the neutron-proton
and proton-proton potentials were the same and if the neutron and
proton density distribution in the target nucleus were also the
same. For the valence particle the sum, $\sum_kV(r_j-r_k)$, in this
case, would reduce to the mean-field potential for this particle and
we recover the one body approximation.

However, because of the Coulomb potential, the proton-proton and
proton-neutron potentials are very different. In contrast to
eq.~\ref{eq6}, the potential in eq.~\ref{eq8} is short-ranged and does
not have a Coulomb tail; the integrals will be restricted to the range
of the nuclear potential. This difference suggests that one-body
models cannot be used to accurately calculate the asymptotic
normalization. They should, however, still give the correct energy
dependence for low energies since they have the correct singularity
structure and asymptotic $r$ dependence.

The cancellation of Coulomb potential contributions in eq.~\ref{eq8}
is unique to dipole transitions. If $e_{i}$ were proportional to the
particle mass, $m_{i},$ then the dipole operator would be just the
center-of -mass coordinate which can not change the internal structure
of the nucleus and the total cross-section would be zero.

Eq.~\ref{eq8} may lead to significant gains in computing $E_{1}$
transitions since the range of the integral has decreased to less then
10~fm from over 100~fm.  Part of gain is lost since the new expression
involve two-body rather then one-body matrix elements. There is
another less obvious drawback. For the $^{7}$Be$(p,\gamma )^{8}$B
reaction the scattering wave function has a node in the nuclear
interior. This leads to large negative and positive contributions to
the integrals in eqs.~\ref{eq6} and \ref{eq8}. There is a large
cancellation between the negative and positive contributions. This
will make the numerical calculation of these matrix elements less
stable.

We now illustrate the points made above by using a single particle
model. We take the potential to have a Woods-Saxon
shape\cite{jennings} and include the Coulomb potential. In the direct
calculation, i.e. using eq.~\ref{eq4}, of the s-wave contribution to
the S-factor the upper bound on the integral must be 300fm for a 2\%
accuracy. The accuracy is determined by first using a large upper
limit on the integral and then reducing it until a 2\% error is
observed. The calculation of the matrix element was then done using
eq.~\ref{eq6}. The results, as expected, agree within numerical
errors to those from eq.~\ref{eq4}.  Using eq.~\ref{eq6} the
upper limit can be reduced to 50fm while still maintaining 2\%
accuracy. This is still quite a large radius and is due to the long
range nature of the Coulomb potential and the previously noted
cancellation in the integral.

In calculations with a two body potential the Coulomb contribution
cancels so we have used eq.~\ref{eq6} to carry out a calculation but
only included the nuclear potential in the $V$. In this case it is
only necessary to integrate to 7fm in order to have 2\% accuracy. This
is the range we would expect from the size of the nucleus. Dropping
the Coulomb contribution from $V$ in eq.~\ref{eq6} also causes the
integral to increases by more than a factor of 2.  This sensitivity to
the presence of the Coulomb potential in eq.~\ref{eq6} indicates that
a single particle model cannot be used to determine the absolute value
of the S-factor; or at a minimum the underlying two-body nature of the
interaction must be explicitly taken into account in calculations that
rely on single-particle wave-functions.

The proton distribution in $^{7}$Be probably has a larger radius than
the neutron distribution. This is due to the Coulomb potential and the
presence of an extra proton. Taking this into account by reducing the
potential radius in eq.~\ref{eq6} (remember that for a two-body
potential only the neutron-proton potential comes in) we find extreme
sensitivity of the S-factor matrix element to the exact radius
used. Reducing the potential radius by just 0.12fm causes the integral
to vanish. In varying the radius of the potential we have kept the
same wave-function since the distortion is due to both the protons and
neutrons. We note that in ref.~\cite{muk2} the result was also
strongly sensitive to the potential used. The sensitivity to
theoretical input may not be just a shortcoming of the technique but
may have a deeper origin indicating that the asymptotic normalization
may, in the end, have to be determined experimentally.

Eq.~\ref{eq8} was derived on the assumption that the two-body
potential had a simple local form. Non-localities or charge exchange
potentials will change this. Consider a general many body Hamiltonian,
$H=\sum _{i}\frac{\nabla _{i}^{2}}{2m_{i}}+\frac{1}{2} \sum
_{i,j}V_{i,j}$.  In eq.~\ref{eq5}, we take the commutator of the
Hamiltonian with the dipole operator. This will, in general, give two
terms. One comes from the kinetic energy and is just $\sum
_{i}\frac{e_{i}\nabla _{i}}{m_{i}}$. The second contribution is from
the commutator of the potential with the dipole operator.  Since the
second term is already short ranged we take the second commutator only
with the first term This gives:
\begin{eqnarray}
M &=& -\frac{\langle\psi _{i}(k)|\sum _{j}\frac{e_{j}\nabla
 _{j}}{m_{j}}|\psi _{f}\rangle}{E_{k}-E_{f}}+\frac{\langle\psi
 _{i}(k)|[V,\sum _{j}e_{j}r_{j}]|\psi
 _{f}\rangle}{E_{k}-E_{f}}\label{eq9a} \\ &=& \frac{\langle\psi
 _{i}(k)|[V,\sum _{j}\frac{e_{j}\nabla _{j}}{m_{j}}]|\psi
 _{f}\rangle}{\left( E_{k}-E_{f}\right) ^{2}}+\frac{\langle\psi
 _{i}(k)|[V,\sum _{j}e_{j}r_{j}]|\psi
 _{f}\rangle}{E_{k}-E_{f}}\label{eq9}
\end{eqnarray}
This equation will be correct for any two-body potential. Since the
Coulomb potential is local it will not contribute to the second
commutator and, as we saw above, cancels in the first commutator. All
the remaining contributions in either term will be of the same range
as the nuclear potential. Thus we see that even in the more general
case the relevant expression can be reduced to matrix elements of
short ranged potentials.

Strictly speaking, neither eq.~\ref{eq3b} nor \ref{eq9} prove that the
matrix elements on the left hand sides have singularities since the
matrix elements on the right hand side may go to zero. For capture
from the d-wave in the $^{7}$Be$(p,\gamma)^{8}$B reaction the Coulomb
wave function has a zero near the pole location so the pole is
effectively canceled and the d-wave contribution to the s-factor is
almost linear in the low energy region\cite{jennings}. In other
derivations\cite{jennings} of the existence of the pole such
accidental zeros must also be checked for. With the exception of such
special cases the procedures outlined in this paper provide the
cleanest and simplest method to show the existence of the pole in the
S-factor and to isolate its residue.

To summarize we have shown that the asymptotic normalization needed to
calculated $S_{17}$ can be obtained from an integral over the nuclear
interior. This relies on the intimate connection between the
wave-function in the nuclear interior and the asymptotic normalization
that follows directly from the bound-state Lippmann-Schwinger
equation.  The matrix element needed in the calculation for $S_{17}$
at any energy can also be reduced to a form that involves only
integrals over the nuclear interior.  These expressions make the
singularity structure of the S-factor explicit and may simplify
calculating $S_{17}$.

ACKNOWLEDGEMENT: The author thanks S.~Gurvitz, S.~Karataglidis,
S.~Scherer and H.W.~Fearing for useful discussions. J.~Escher is
thanked for carefully reading the manuscript and useful
discussion. The Natural Sciences and Engineering Sciences Research
Council is thanked for financial support.

\end{document}